\documentclass[journal,a4paper,10pt]{IEEEtran}
\usepackage{amsmath,amssymb,amsfonts} % Typical maths resource packages
\usepackage{graphicx}                 % Packages to allow inclusion of graphics
\usepackage{fontenc}
\usepackage{multirow} %% Package for multiple rows in a table
\usepackage{theorem}
\usepackage{array}

\usepackage{color,cite}
\usepackage{algorithmic}
\usepackage{algorithm}
\usepackage{epstopdf}
\usepackage{esint}
\usepackage{mathrsfs}
\usepackage[english]{babel}
\usepackage[table,xcdraw]{xcolor}
\usepackage{booktabs}
\usepackage{multirow,makecell}

\usepackage{array}
\newcolumntype{P}[1]{>{\centering\arraybackslash}p{#1}}
\newcolumntype{M}[1]{>{\centering\arraybackslash}m{#1}}

\begin{document}

\title{Dense Small Satellite Networks for Modern Terrestrial Communication Systems: Benefits, Infrastructure, and Technologies}

\author{{Naveed UL Hassan,~\IEEEmembership{Senior Member,~IEEE,} Chongwen Huang,~\IEEEmembership{Member,~IEEE,} Chau Yuen,~\IEEEmembership{Senior Member,~IEEE,} Ayaz Ahmad,~\IEEEmembership{Senior Member,~IEEE,} and Yan Zhang,~\IEEEmembership{Fellow,~IEEE}}
\thanks{N. U. Hassan is with the Department of Electrical Engineering at Lahore University of Management Sciences (LUMS), Lahore, Pakistan 54792. (Email: naveed.hassan@lums.edu.pk).}
\thanks{C. Huang and C. Yuen are with the Engineering Product Development at the Singapore University of Technology and Design (SUTD), 8 Somapah Road, Singapore 487372. (Email: chongwen\_huang@mymail.sutd.edu.sg, yuenchau@sutd.edu.sg).}
\thanks{A. Ahmad with the Electrical and Computer Engineering Department, COMSATS University Islamabad - Wah Campus, Wah Cantt., Pakistan, 47040 (Email: ayaz.uet@gmail.com).}
\thanks{Y. Zhang is with Simula Research Laboratory, Norway; and Department of Informatics, University of Oslo, Gaustadalleen 23, Oslo, Norway (Email: yanzhang@simula.no).}

\thanks{This research was supported by LUMS, Faculty Initiative Fund (FIF), in part by the SUTD-MIT International Design Centre (idc: idc.sutd.edu.sg), and the National Natural Science Foundation of China under Grant 61941102.}
}
\maketitle

\begin{abstract}
Dense small satellite networks (DSSN) in low earth orbits (LEO) can benefit several mobile terrestrial communication systems (MTCS). However, the potential benefits can only be achieved through careful consideration of DSSN infrastructure and identification of suitable DSSN technologies. In this paper, we discuss several components of DSSN infrastructure including satellite formations, orbital paths, inter-satellite communication (ISC) links, and communication architectures for data delivery from source to destination. We also review important technologies for DSSN as well as the challenges involved in the use of these technologies in DSSN. Several open research directions to enhance the benefits of DSSN for MTCS are also identified in the paper. A case study showing the integration benefits of DSSN in MTCS is also included.

\end{abstract}

\section{Introduction}
Communication satellites are in service since 1960s. Over the years, satellite communication networks provided niche services and acted as relays for long distance broadcasts and for connectivity in remote and harsh environments. Satellites also provide positioning and timing synchronization services for various mobile terrestrial communication systems (MTCS). However, satellite communication systems always remained expensive, proprietary, and have been largely developed independently of various MTCS.

\begin{figure}[tpb]
\centering
 \includegraphics[width=7.0cm]{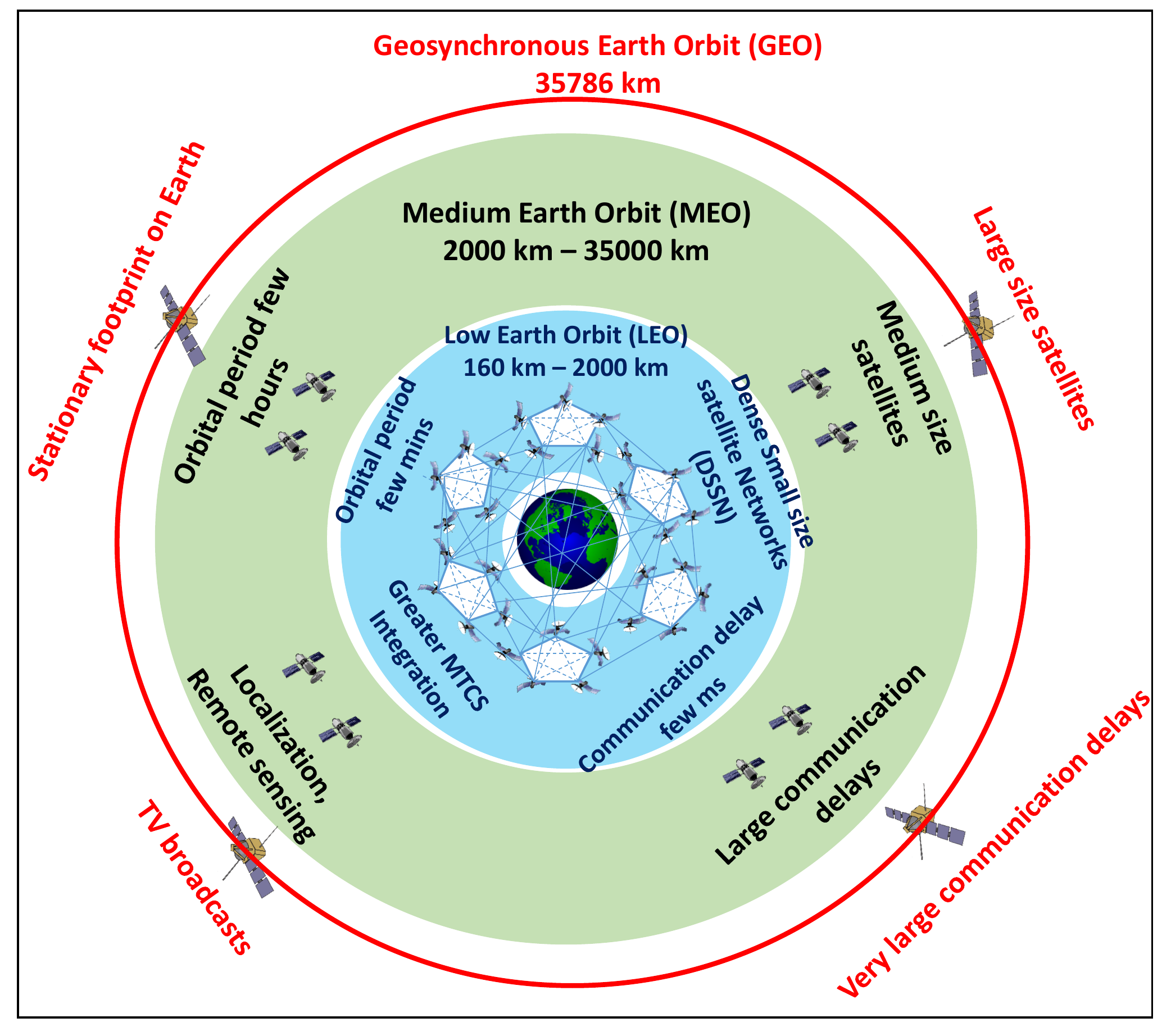}
\caption{GEO, MEO and LEO satellites.}
\label{fig:geomeoleo}
%\vspace{-0.22in}
\end{figure}

In recent years, with growing demands for high data rate applications, massive connectivity, universal internet access, Internet of Things (IoT) and wireless sensor networks (WSNs), there is a renewed focus and an ever increasing interest in small low earth orbit (LEO) satellites. LEO satellites typically weigh less than 500kg \cite{di2019ultra}, and the development cost, launch cost, and propagation delays of these satellites are significantly less than the traditional large-sized medium earth orbit (MEO) and geosynchronous Earth orbit (GEO) satellites. However, capabilities and resources of a single small satellite are relatively limited. Therefore, several companies, such as, SpaceX, OneWeb, Kepler, and SPUTNIX have announced plans to develop dense small satellite networks (DSSN) comprising of thousands of satellites as shown in Figure \ref{fig:geomeoleo}. The initial setup cost of DSSN is huge, low-earth orbits would also impact the expected lifetime of satellites due to ionizing effects of particles trapped in Earth's magnetosphere, collision avoidance would require careful orbital planning, and de-orbiting of non-functional satellites can also become challenging with time. Nonetheless, DSSNs will put huge amount of communication resources in space that could benefit various MTCS including cellular systems, IoT / WSN, vehicular / drone networks, and smart infrastructure such as smart power grids.

The rest of the paper is organized as follows. In Section~\ref{sat:app} we discuss the potential benefits of DSSN for MTCS. In Section~\ref{sat:arch} we discuss important components of DSSN infrastructure and describe DSSN system architectures from physical and link layer aspects. DSSN technologies and challenges are discussed in Section~\ref{sat:tech}. A case study on the integration benefits of DSSN in MTCS is provided in Section~\ref{sat:example}. The paper is concluded in Section~\ref{sat:conc}.

\section{Potential DSSN Benefits for MTCS}
\label{sat:app}
\begin{figure}[tpb]
\centering
 \includegraphics[width=8.0cm]{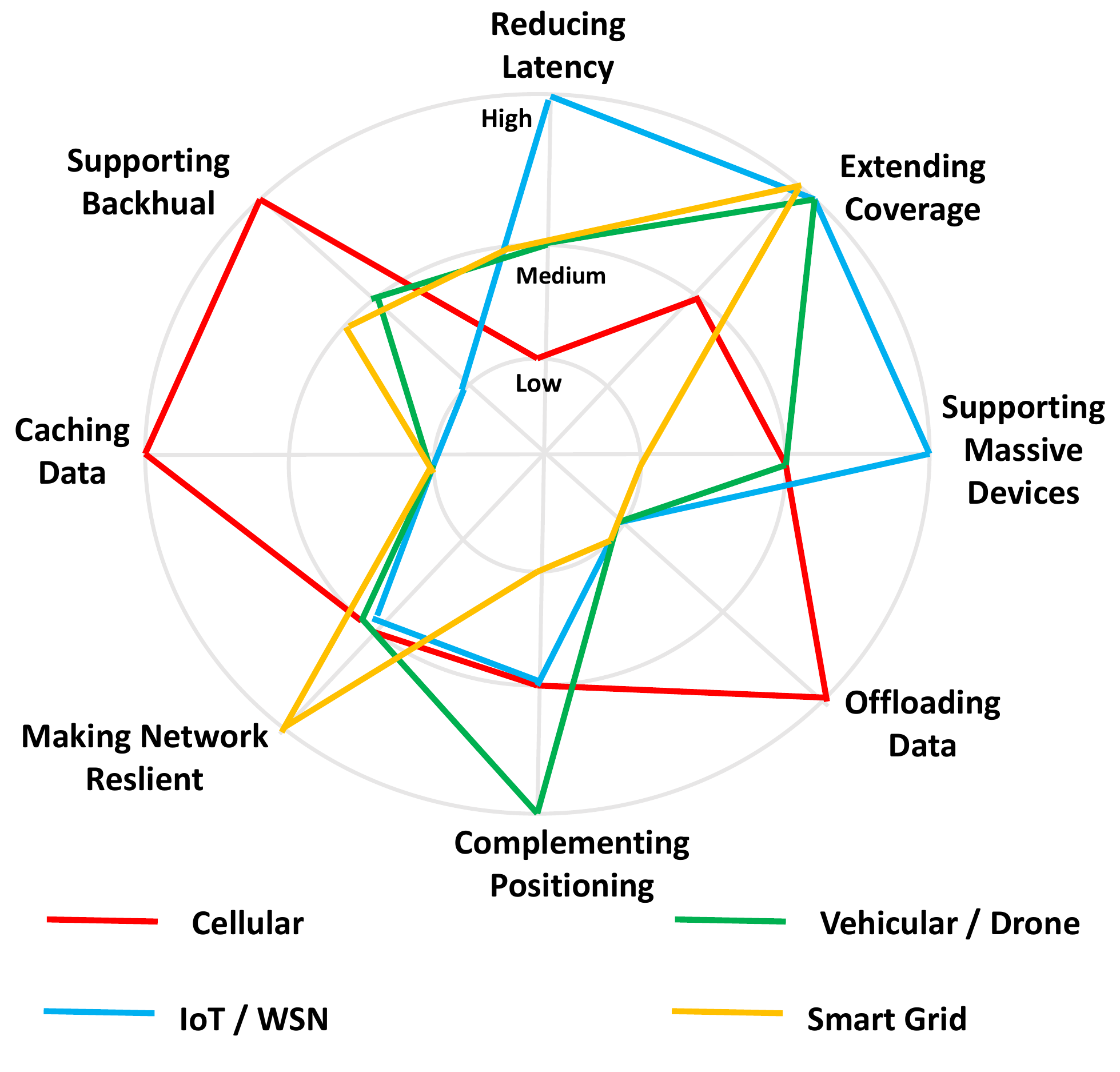}
\caption{Potential DSSN benefits for MTCS.}
\label{fig:benefits}
%\vspace{-0.22in}
\end{figure}

In this section we discuss some potential benefits of DSSN for MTCS.
These benefits are also depicted through Figure~\ref{fig:benefits}.

\subsection{Extending Coverage} The most significant benefit of DSSN is its wide area coverage, which can benefit mobile users, sensors, and vehicles in remote, challenging and under-developed regions. Extended coverage through DSSN can also benefit several smart grid applications such as, remote monitoring of offshore renewable wind farms, distributed sub-stations, and transmission \& distribution networks. It can also provide autonomous vessel support, which cannot be easily done with today's telecom networks.
A single LEO satellite can approximately cover 1~million km$^2$ area on Earth.
Due to huge coverage area of each satellite, spot beam coverage and hybrid wide-spot beam coverage schemes are typically employed \cite{su2019broadband}. In spot beam coverage scheme, each satellite provides multiple spot beams whose footprint on Earth also moves along with the satellite trajectory. On the other hand, in hybrid wide-spot beam scheme, each satellite provides a low-power wide beam for the whole service area and high power spot beams are digitally formed. The spot beams in hybrid wide-spot beam scheme are steered to the terrestrial users in order to achieve a fixed footprint on Earth during the satellite trajectory. In addition, several satellites can also coordinate with each other to provide focused and narrow spot beams.

\subsection{Reducing Latency} There is an increased focus on achieving ultra-low-latency in several MTCS. Due to lower altitude orbits, round trip time of signals between a LEO satellite and a ground terminal is only few ms (30ms for OneWeb system and 10-15ms for SpaceX Starlink system) \cite{su2019broadband}. This latency is enough to fulfill the requirements of many IoT, smart grid, and vehicular communication applications. The objective of cellular systems such as, 5G and beyond is to achieve 1ms latency, which cannot be directly attained with the help of DSSN. However, DSSN may indirectly facilitate 5G networks in reducing latency by providing alternate backhaul and data caching options. It is also important to note that as compared to free space, the speed of light is 30-40\% slower in fiber . Therefore, for long distance communication, DSSN has the potential to provide lower latency as compared to any terrestrial network of comparable length. However, due to relative movement between satellites, latency may vary, communication links may become unstable and handovers may increase.

\subsection{Supporting Massive Devices} The number of mobile users, IoT devices, and autonomous vehicles is increasing at an unprecedented pace. With large amount of resources in space, one great benefit of DSSN is that it can help support an order of magnitude more devices. DSSN is also more cost-efficient for IoT devices deployed in deserts, forest, oceans and other challenging areas. Due to the mobility of LEO satellites, devices can communicate with LEO satellites at different elevation angles thus providing more tolerance to terrestrial obstacles. Path loss due to lower orbital altitudes of LEO satellites is also smaller, which can help support more low-powered devices. However, key challenges include, interoperability issues, efficient medium access protocols, and optimization of available resources.

\subsection{Offloading Data} DSSN can be integrated in cellular networks by creating LEO small cells (LSC) \cite{di2019ultra}. In LSC, a terrestrial node with satellite connectivity acts as a base station (BS) and serves multiple cellular users. LSCs can coexist with traditional small cells or macro BS. In this arrangement, some data traffic can be offloaded from the terrestrial network.
LSCs maybe set up in areas with heavy traffic spikes or in rural areas with no terrestrial communication networks. LSCs can also be dynamically created in geographical regions experiencing demand spikes with the help of Unmanned Aerial Vehicle (UAV) and autonomous vehicles. Such setup is more useful in dense urban environments with huge data rate demands or in highly stressed data networks, and would require advance traffic prediction algorithms. Such integration opportunities can also be exploited in any MTCS supporting a large number of devices with limited resources and can further enhance network resilience.

\subsection{Complementing Positioning} The knowledge of exact location of user or device is often required to provide useful location based services. In cellular systems, prediction of user or device mobility can facilitate content caching. In vehicular communication systems, positioning and navigation are essential to avoid accidents. In IoT and smart grid networks, positioning may also be needed for asset tracking. DSSN can complement GPS by providing an alternate way to determine the location of users and devices whenever GPS signals are out of range. The high mobility of LEO satellites results in an extremely large Doppler shift and Doppler frequency rate of change. With some information about satellite orbits and positions, ground receivers can use Doppler measurements for localization. Furthermore, time difference of arrival and frequency difference of arrival measurements can also be used to determine the location of the ground receiver. However, achieving very high positioning accuracy with DSSN is an open research challenge.

\subsection{Making MTCS more Resilient} DSSN can enhance the overall resilience of MTCS. Network congestion and overloading can be avoided with the help of additional and redundant DSSN connections. These connections would also be valuable in case of emergency and disasters scenarios. For critical smart grid infrastructure, resilient communication network is also critical. DSSN is more tolerant to extreme topographies and is also more robust against challenging terrestrial environments.

\subsection{Caching Data} Data caching entails storing popular or frequently accessed content closer to the end users. It is an important method to reduce latency and backhaul congestion in cellular networks. With vast coverage, DSSN can help bring content closer to the end users. To save resources in terrestrial networks, some data can also be cached in DSSN \cite{wu2016two}. Satellites in DSSN also have the ability to multi-cast data and quickly update the cached content at multiple locations. Development of effective caching strategies involving DSSN are interesting research directions. The use of information centric networking paradigm in which the content can be retrieved by its name and every network node has some cache space can also be explored for managing content caching and content delivery in such systems. However, each LEO satellite has limited storage and computing capabilities, therefore, limited content placement opportunities exist, which could also create strong competition among MTCS, and several interesting game theoretic models may be applied to determine the price of data caching in DSSN.

\subsection{Supporting Backhaul} The amount of traffic generated by users and devices is exponentially increasing. Ultra-dense deployment of small cells is a major technique to support huge traffic demand in 5G networks. However, data generated in the these small cells and frequent handover requests due to user mobility can put huge pressure on the backhaul resources. DSSN can help resolve backhaul capacity issues by providing wideband communication links. Multiple LEO satellites also provide more dimensions for backhaul capacity maximization. Backhaul capacity of DSSN links can dynamically vary depending on the satellite orbital paths and the technology used on various communication links. These variations can be exploited to advantage through developing dynamic scheduling and resource optimization algorithms.

\section{DSSN Infrastructure}
\label{sat:arch}
LEO satellites in DSSN circle the Earth at altitudes extending from 160km to 2000km. These satellites move at extremely fast speeds (several thousand km/hr) and typically complete one orbital revolution in few hundred minutes. From any specific point on Earth, a satellite is only available for few minutes during each revolution. The potential DSSN benefits for MTCS can be achieved through carefully planned DSSN infrastructure. In this section, we discuss various components of DSSN infrastructure and summarize our discussion in Table~\ref{sat:arc}.

\begin{table*}[htb]
\centering
\fontsize{8pt}{8pt}\selectfont
\caption{DSSN Infrastructure}
\begin{tabular}{c | M{2.4cm} | M{3.6cm} | M{4.0cm} | M{4.0cm}} \hline
\rowcolor[HTML]{9B9B9B}
DSSN Component & Type & Description & Advantages & Disadvantages \\ \hline
 \multicolumn{1}{c|}
{\multirow{2}{*}{\parbox{2.2cm}{\centering \textbf{Satellite Formations}}}} &  Constellation & All satellites are identical. & Low cost and high redundancy. & Required to fly in well planned orbits.    \\ \cline{2-5}
  & Cluster  & Non-identical but can cooperate. & Individual modules can be replaced instead of whole satellites. & Complicated and high cost design.  \\ \hline \hline
	
\multicolumn{1}{c|}{\multirow{3}{*}{\parbox{2.2cm}{\centering \textbf{Satellite Orbits}}}} & Polar & All the orbital planes pass over Earth poles. & Easy to predict satellite path and high coverage over polar regions. & Low coverage away from poles. \\ \cline{2-5}
& Rosette & Highly inclined orbits to provide greater coverage away from poles. & Minimum five satellites are needed to cover the entire Earth. & Less coverage around poles. \\ \cline{2-5}
& Hybrid   & Mixture of polar and rosette orbits. & High coverage flexibility. & High complexity.   \\ \hline \hline

\multicolumn{1}{c|}{\multirow{3}{*}{\parbox{2cm}{\centering \textbf{ISC Links}}}} & RF & Communication takes place over radio spectrum. & Multiple bands UHF, S, K, Ka, Ku, etc. Several design tradeoffs are possible. & Links are susceptible to interference, large antennas are required for communication over long distances. \\ \cline{2-5}
& OWC  & Free space optical communication in which modulated data is transmitted on unguided channels. Wavelengths are in the range of 500nm to 2000nm. & High directivity, high bandwidth, high security, low power consumption. &  High costs and strict beam alignment challenges. \\ \cline{2-5}

& VLC  & Simple LED lights can be used as transmitters. & Low cost and very low power consumption. & Greater background illumination noise and design of optical filters before photo-detectors could be challenging.  \\ \hline \hline

\multicolumn{1}{c|}{\multirow{4}{*}{\parbox{2cm}{\centering \textbf{SGC/GSC Architecture}}}} & Direct communication with destination & Direct communication with the destination node. & Simple architecture and no requirement of ISC links. & Extremely high worst case latency.  \\ \cline{2-5}

& Communication with the aid of ground infrastructure & Source node transmits data to its nearest ground station which transmits data to the destination node. & Moderate latency and no requirement of ISC links. & Requires reliable terrestrial communication network and gateways in case of different protocols and communication technologies.  \\ \cline{2-5}

&  Communication with the aid of space infrastructure & Data is first routed in space to the satellite closet to the destination node. & Very low latency can be achieved with fast ISC links. & Requires a fully connected space network.  \\ \cline{2-5}

&  Communication with the aid of space \& ground infrastructure & Data can be routed through space or ground infrastructure nodes. & Extremely flexible, and can achieve very low latency. & Complex design, requires ISC links as well as reliable terrestrial communication network.  \\ \bottomrule

\end{tabular}
\label{sat:arc}
\end{table*}

\subsection{Satellite Formations}
Large number of small satellites may be deployed in two different formations.

\begin{enumerate}
	
	\item \textbf{Constellation}: This formation consists of multiple replicas of the same satellite. All the satellites have the same hardware and perform identical functions.
		
	\item \textbf{Cluster}: A cluster is group of non-identical satellites, which can cooperate with each other. Every satellite in the cluster has its specific role.
	
\end{enumerate}

Constellation is the preferred formation choice because it brings simplicity and reduces the manufacturing and deployment costs of DSSN. Terrestrial systems generally undergo rapid technological evolutions, which can potentially limit the lifetime of constellations because replacement of older technology in space remains quite challenging. However, some re-configurability may be achieved through the use of software defined radios (SDR) \cite{maheshwarappa2018reconfigurable}.

\subsection{Satellite Orbits}
Satellites in DSSN can be arranged in one or more orbital planes at different altitudes and inclinations. While designing orbital paths for DSSN, it is more important to focus on those arrangements in which satellite positions can be conveniently predicted. This simplifies the creation of DSSN, handover mechanisms, contact planning, energy harvesting, positioning, and resource allocation. Some classical methods to plan predictable satellite orbits are:

\begin{enumerate}

\item \textbf{Polar}: In this arrangement, all the orbital planes pass over Earth poles. Each orbital plane has one or more satellites. These orbits provide high degree of coverage over the poles.

\item \textbf{Rosette}: In this arrangement, orbital paths are highly inclined relative to the equator in order to provide high coverage away from the poles. In rosette (flower like) constellation, minimum five satellites are needed for continuous worldwide visibility.

\item \textbf{Hybrid}: In hybrid design, a mixture of polar and rosette arrangements may be used to achieve different coverage in different regions.

\end{enumerate}

In the literature, satellite orbital planning has been studied for relatively small number of satellites with primary focus on providing worldwide coverage. However, with thousands of satellites in DSSN providing support to various MTCS with diverse objectives such as, latency minimization, data rate maximization, and power consumption minimization, orbital planning problem becomes more challenging.

\subsection{Inter-Satellite Communication} A network of satellites can only be developed in space through the creation of ISC links \cite{radhakrishnan2016survey}. Such links enable communication and cooperation among satellites for data routing, throughput maximization, latency minimization, and seamless coverage. Below we discuss some communication technologies for ISC links. These technologies can also be used for satellite-to-ground communication (SGC) and ground-to-satellite communication (GSC) links.

\begin{enumerate}

\item \textbf{Radio Frequency (RF) links}: There are several RF bands that can be used for ISC. The choice of RF band, modulation and coding scheme, error detection and correction mechanism, and antenna size in the design of ISC links primarily depends on the distribution of satellites, sources of RF interference, and the integration objectives of DSSN in MTCS.

\item \textbf{Optical Wireless Communication (OWC) links}: On OWC links, modulated light signals are transmitted in free space using lasers, which provide very high directive gains, high data rates, high security, and low power consumption. However, OWC links suffer from high costs, strict alignment challenges of optical beams, and high background illumination noise. Laser wavelengths considered for ISC links are generally in the range of 500nm to 2000nm. Longer wavelengths should be preferred because they cause reduction in solar background and solar scattering.

\item \textbf{Visible Light Communication (VLC) links}:
In this method, LED lights are used to transmit data in visible light spectrum (350nm - 750nm). However, background illumination due to Sun in LEO orbits is very high (around 580 W/m$^2$) which can easily drown any useful communication signal. To achieve meaningful communication, optical filters have to be placed before the photo-detectors. These filters exploit the so called Fraunhofer lines, which are the frequencies in VL spectrum that get absorbed by chemical elements present in the Sun. The use of VLC is only recommended for short to medium range links.

\end{enumerate}

Creation of ISC links in DSSN is quite challenging due to satellite weight and power limitations.
When large number of satellites in DSSN are in the coverage range of each other and when the network topology is dynamically changing, acquiring and maintaining a robust communication link without interference also becomes difficult. High gain, multi-band, multi-functional, and multi-beam, smart steerable satellite antennas can overcome some of these challenges \cite{gao2018advanced,zheng2019time}. Joint optimization problems involving orbital path planning and ISC link design can also be formulated according to the requirements of MTCS.

\subsection{SGC / GSC System Architectures}
\begin{figure}[tpb]
\centering
 \includegraphics[width=0.49\textwidth]{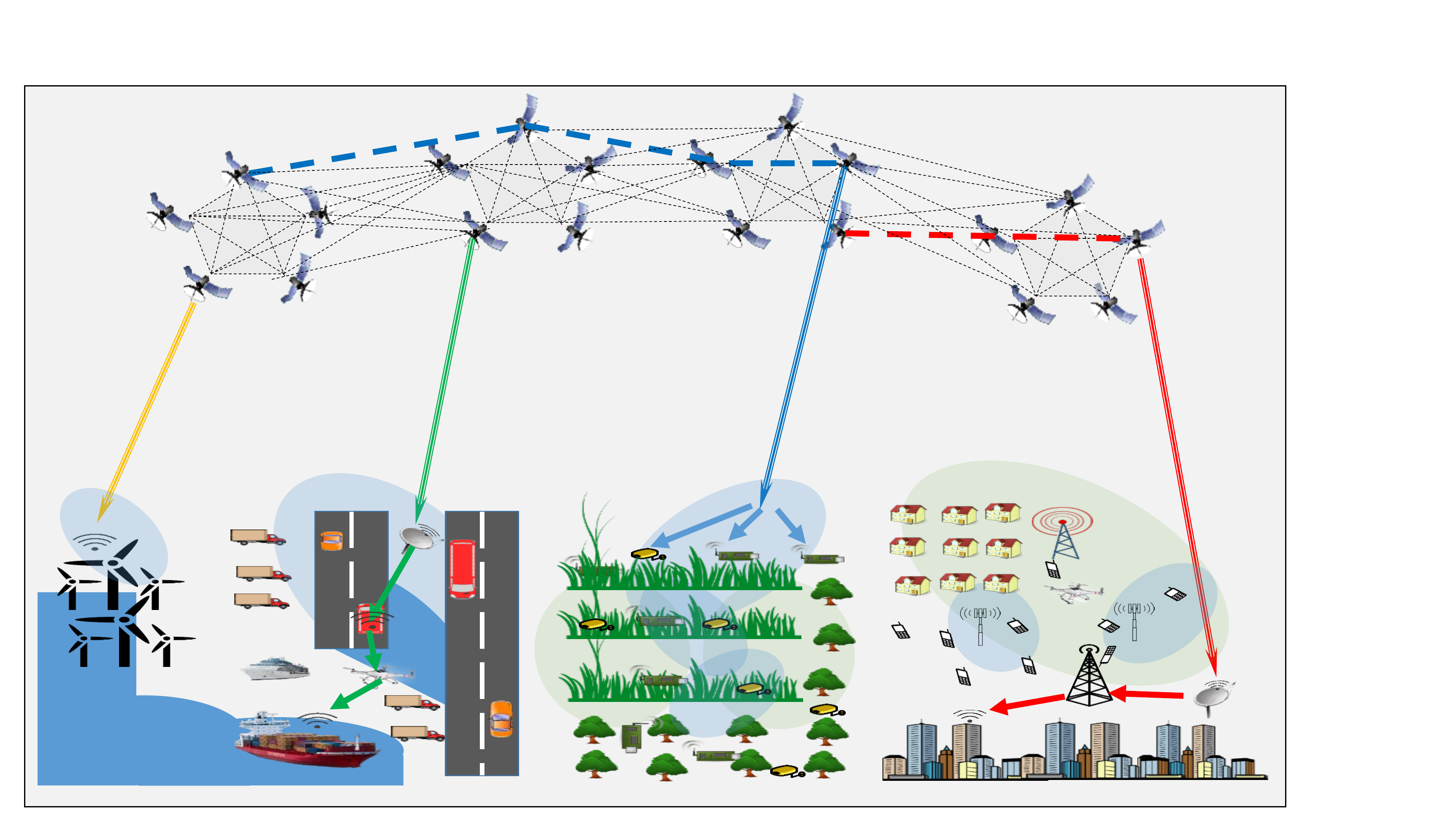}
\caption{SGC system architectures in various MTCS Scenarios.}
\label{fig:sgc}
\end{figure}

We discuss four different SGC / GSC system architectures. To facilitate the understanding and design of these communication architectures, we assume source node is located in DSSN and destination node is located on ground. Please note that these architectures only describe the physical and data link layer aspects. These communication system architectures are also described with the help of Figure~\ref{fig:sgc}.

\begin{enumerate}
	\item \textbf{Direct communication with destination}: In this communication system architecture, source node in DSSN directly communicates with the destination node on ground. LEO satellites are non-stationary and in this case source node can only communicate when the destination node is in its coverage area. This is a simple architecture and it can be used when DSSN does not support ISC links and there is no supporting communication infrastructure on ground.
	
	\item \textbf{Communication with the aid of ground infrastructure}: In this communication system architecture, when the source node has data for the destination node it immediately transmits this data to its nearest ground station. Once the data is at the ground station, it is transmitted to the destination node using traditional terrestrial communication networks. This method is also helpful when DSSN has no ISC links. Latency would primarily depend on the performance of the terrestrial network.
	
	\item \textbf{Communication with the aid of space infrastructure}: This communication system architecture is only possible in DSSN with ISC links. In this method, data is first routed from source node to the satellite closest to the destination node. The intermediate DSSN satellite then transmits data to the destination node. This approach has the potential to drastically reduce latency.
	
	\item \textbf{Communication with the aid of space \& ground infrastructure}: This is the most flexible communication system architecture. In this architecture, the source node transmits information through intermediate satellites and ground infrastructure for faster delivery of data. This method can make the best use of all the available resources but it requires the availability of DSSN with ISC links and ground infrastructure.

\end{enumerate}

In communication system architectures 1 and 2, there are no ISC links and LEO satellites require physical layer protocols to establish a link with the gateways/nodes on the ground. On the other hand, in communication system architectures 3 and 4, satellites are equipped with ISC links and therefore, in addition to physical layer protocols, medium access protocols (MAC) and networking protocols are also required on each satellite. In all these architectures, appropriate gateways comprising of antennas, baseband processing units, router and core network entities are also required on the ground \cite{su2019broadband}. With ISC links, the number of ground nodes and gateways can be drastically reduced. However, creating a network in space has its own challenges due to limited on-board power supply and fast orbital speeds.

\section{DSSN technologies and challenges}
\label{sat:tech}
In this section, we discuss important technologies that can help achieve DSSN benefits for MTCS. Different technologies, associated challenges, and possible solutions are also provided in Table~\ref{sat:techno}.

\begin{table*}[htb]
\centering
\fontsize{9pt}{9pt}\selectfont
\caption{DSSN Technologies, Challenges and Solutions}
\begin{tabular}{|c|M{5.5cm}|M{7.5cm}|M{1.5cm}|} \hline
\rowcolor[HTML]{9B9B9B}
\textbf{DSSN Technology} & \textbf{Challenges}  & \textbf{Possible Solutions} & \textbf{Surveyed Papers} \\ \hline

\multirow{2}{*}{\parbox{2cm}{\centering \textbf{Smart Steerable Satellite Antenna}}}

& {\parbox{5.5cm}{\centering High gain, light weight and low power consumption antenna designs}} & {\parbox{7.5cm}{\centering Null scanning retro-directive antenna array, Bull's Eye antenna with multiple annular rings, integration of low-loss tunable materials}} & \cite{gao2018advanced} \\\cline{2-4}

& {\parbox{5.5cm}{\centering Multi-band, multi-beam and steerable antenna designs}} & {\parbox{7.5cm}{\centering Beam scanning antennas using slot active frequency selective surfaces, use of optical switches with activation techniques}}  & \cite{zheng2019time}  \\ \hline

 \multirow{2}{*}{\parbox{2cm}{\centering \textbf{Multiple Access Techniques}}}
& {\parbox{5.5cm}{\centering Diverse satellite network architectures}}  & {\parbox{7.5cm}{\centering Conventional, cooperative, cognitive NOMA techniques can be used according to the network architecture and MTCS characteristics}}  & \cite{zhang2020user} \\\cline{2-4}

& {\parbox{5.5cm}{\centering Additional interference due to satellite beam widening and other sources}} & {\parbox{7.5cm}{\centering Overlay coding in multi-beam satellite and optimized user pairing strategies}} & \cite{beigi2018interference} \\ \hline

\multirow{2}{*}{\parbox{2cm}{\centering \textbf{Energy harvesting and optimization}}}
& {\parbox{5.5cm}{\centering Variable and limited harvested energy by satellite with changing traffic demand}} & {\parbox{7.5cm}{\centering Contact plan optimization and novel energy optimization algorithms}} &  \cite{fraire2018battery}  \\\cline{2-4}

& {\parbox{5.5cm}{\centering Optimal use of satellite battery for lifetime maximization}} & {\parbox{7.5cm}{\centering Joint consideration of battery lifetime maximization, energy efficiency and Quality of Service (QoS) requirements of path length and the maximum link utilization ratio}} & \cite{yang2016towards} \\ \hline

{\parbox{2cm}{\centering \textbf{Routing and Networking}}}
& {\parbox{5.5cm}{\centering Changing satellite topologies due to motion}} & {\parbox{7.5cm}{\centering Dynamic routing algorithms with various routing metrics according to changing requirements}} & \cite{radhakrishnan2016survey} \\ \hline % \\\cline{2-4}

\multirow{2}{*}{\parbox{2cm}{\centering \textbf{Re-configurability}}}
& {\parbox{5.5cm}{\centering Replacement of older technology on satellites due to rapid evolution of terrestrial technologies}} & {\parbox{7.5cm}{\centering Greater re-configurability options with the help of SDR, combining low-cost SDR hardware with open-source software tools for DSSN}}  &  \cite{maheshwarappa2018reconfigurable} \\\cline{2-4}

& {\parbox{5.5cm}{\centering Spectrum scarcity \& efficient spectrum utilization in DSSN}} & {\parbox{7.5cm}{\centering Use of CR technologies by considering satellite users as secondary users, which then coexist with licensed spectrum users and exploit spectrum in interweaving, overlay and underlay fashion }} &   \cite{ferreira2018multiobjective}
\\ \hline

\multirow{2}{*}{\parbox{2cm}{\centering \textbf{Data Caching}}}
& {\parbox{5.5cm}{\centering Selection of satellites and appropriate content for data caching}} & {\parbox{7.5cm}{\centering Content popularity prediction and distributed content management systems based on various objectives, two layer content caching schemes where content can be cached in ground nodes as well as satellite nodes}} &  \cite{wu2016two} \\\cline{2-4}

& {\parbox{5.5cm}{\centering Limited cache space in small satellites}} & {\parbox{7.5cm}{\centering Multi-layer satellite caching problems where larger GEO satellites can offer its space to competing LEO satellites }} & \cite{wang2019load} \\ \hline

\multirow{2}{*}{\parbox{2cm}{\centering \textbf{Resource Optimization}}}
& {\parbox{5.5cm}{\centering Complex optimization problems with diverse objectives and constraints}} & {\parbox{7.5cm}{\centering Game Theory, control theory, machine learning, neural networks, and reinforcement learning techniques for problem solving}}  &
\cite{qiu2019deep}
\\\cline{2-4}

& {\parbox{5.5cm}{\centering Lack of diversity on ISC, SGC and GSC links due to direct signal propagation}} & {\parbox{7.5cm}{\centering Multi-cast and multi-group beamforming design over large geographical regions, exploitation of atmospheric conditions and rain attenuation}} & \cite{christopoulos2015multicast}
\\\hline

\end{tabular}
\label{sat:techno}
\end{table*}

\subsection{Smart Steerable Satellite Antenna Designs}
Changing distances and angles between satellites necessitate smart steerable DSSN antennas. Due to relatively large propagation distances, these antennas are also required to have high-gain and ultra-wide bandwidth. Moreover, due to extremely large footprint of a satellite on Earth, satellite antennas should have the capability to produce multiple independent narrow spot beams.

Researchers at the University of Hawaii have developed extremely lightweight (186g) and low power null scanning retro-directive antenna array for ISL links. Low-power and low-cost, multi-beam Bull's Eye antenna with multiple annular rings is also being developed for extremely small satellites. Some low-loss tunable materials such as liquid crystals, ferroelectric thin films, and piezoelectric materials may be used to create multi-beam steerable satellite antennas \cite{gao2018advanced}. For OWC links, steerable satellite antennas may be developed by including optical switching techniques such as, wavelength switching (WS), time slice switching (TSS), and electronic packet switching (EPS), as well as their activation techniques into the design \cite{zheng2019time}. Another promising research direction could be the inclusion of reconfigurable intelligent surface (RIS) on satellites. RIS can be used to passively steer the radio signals and provide additional control on the challenging space environment. However, the integration and cooperation with RIS are not yet explored in literature.

\subsection{Multiple Access in DSSN}
Traditional satellite systems employ orthogonal multiple access (OMA) techniques that allow devices exclusive access to different resource elements (frequency, time slot, spreading code) thereby limiting the full reuse potential of all the available resources. On the other hand, non-orthogonal multiple access (NOMA) techniques allow the simultaneous use of single resource element by multiple devices. With its large coverage area, LEO satellites can serve massive number of devices by employing multi-beam techniques. Full frequency reuse on multiple beams in multi-beam satellites with the help of NOMA techniques can provide high data rates. However, the use of NOMA in satellite communication systems is not straightforward due to huge distances between user terminals in satellite beams. Therefore, user grouping, power allocation for each beam, and channel estimation for successive interference cancellation become more challenging.

The feasibility of NOMA in DSSN is analyzed in \cite{zhang2020user} and the application of conventional, cognitive and cooperative NOMA schemes in different SGC and GSC communication architectures is discussed. The application of NOMA in multi-beam satellite systems has also been investigated in \cite{beigi2018interference}. Overlay coding scheme in which the beams cooperate and the strongest co-channel interference acts as an additional source of information is employed based on optimized user pair strategies.

\subsection{Energy Harvesting \& Optimization in DSSN}
DSSN satellites are generally powered by solar panels. The amount of harvested energy varies according to satellite orbital path and the position of its solar panels relative to the Sun. However, unlike on Earth, solar power fluctuations can be easily predicted when orbital paths of satellites in DSSN are predictable. On the other hand, traffic generated on SGC, GSC, and ISC links is random.

In this context, several interesting energy optimization problems arise. In DSSN, joint energy optimization problems over the entire network or a subset of satellites can be formulated. In \cite{fraire2018battery} battery aware contact plan is designed for DSSN and authors consider detailed battery models and ISC and SGC link budget in their analysis.
Battery lifetime is impacted by frequent charging and discharging cycles, which then impacts the overall performance and operational life of DSSN. In \cite{yang2016towards}, energy efficient DSSN routing problem is formulated with the objective of maximizing the battery lifetime. An algorithm is developed which jointly considers energy efficiency and Quality of Service requirements of path length and the maximum link utilization ratio.

\subsection{Routing and Networking in DSSN}
Routing algorithms can be designed for DSSN based on different routing metrics. Handover optimized routing algorithm uses connection matrix to identify the presence of ISC links, bandwidth-delay routing algorithm considers delay and bandwidth as the routing matrix, destruction resistant routing algorithm enhances the survivability of the network by considering link states as routing metric, Stiener-free routing can support large number of satellites, distributed multi-path routing provides better latency and can track the changing topology of DSSN, while dynamic routing algorithm based on mobile ad-hoc network can provide high autonomy and limited overhead. A comprehensive survey of networking issues in small satellites is available in \cite{radhakrishnan2016survey}.

\subsection{Re-configurability in DSSN} Replacing old technology in DSSN constellations is not easy and in this context, DSSN systems can be made more reconfigurable and adaptable with the help of SDR and cognitive radio (CR) technologies. However, the development of easily reconfigurable SDR based DSSN systems that can process hybrid signals, while maintaining link performance and reliability is a challenging task \cite{maheshwarappa2018reconfigurable}. Novel SDR architectures, hardware and software technologies are required. Moreover, combining existing available low-cost SDR hardware and open-source software tools to achieve greater flexibility in DSSN can also be investigated.

CR technologies can also be employed in DSSN for efficient spectrum exploitation. However, huge coverage areas of satellites and multiple beams in different geographical regions can pose challenges in the spatial reuse of available spectrum in different MTCS. On-board processing capabilities on satellites are also limited and therefore spectrum sensing from space is also challenging. A multi-objective reinforcement learning technique for cognitive satellite communication, which considers radio frequency, orbital paths, atmospheric conditions, space weather conditions, and available satellite memory is proposed in \cite{ferreira2018multiobjective}. The proposed artificial intelligence (AI) technique provides more effective control and spends less time on the estimation of radio parameters that often results in low performance.

\subsection{Data Caching in DSSN}
Due to large area of coverage, distributed DSSN based content management systems can be designed in order to fetch and store popular content. User profiles can also be considered while selecting DSSN satellites for content caching. Some authors have also considered the problem of content placement in DSSN to minimize delay. A two layer content caching scheme is proposed in \cite{wu2016two} where first cache layer is deployed in ground nodes and second cache layer in DSSN. This joint caching optimization technique minimizes satellite bandwidth consumption for content delivery.

In \cite{wang2019load} a multi-layer satellite system comprising of GEO and LEO satellites is considered and the limited cache storage space of GEO satellites is considered as a resource. LEO satellites compete for cache space and Stacklerberg game is developed for load balancing. Machine learning, AI, and reinforcement learning techniques can be used for content prediction and user mobility pattern prediction for the development of efficient data caching algorithms involving DSSN.

\subsection{Resource Optimization in DSSN} Resources in DSSN include, time, spectrum, antenna, satellite beams, orbital planes, and power. When DSSN resources are combined with MTCS and UAV resources, several interesting resource optimization opportunities arise. However, at the same time, the problem also becomes more complex due to very large resource pool, interference issues, and channel state information (CSI) availability challenges.
Resource optimization problems to support different MTCS and UAVs can be formulated with different objectives such as, capacity maximization, power minimization, latency reduction, quality enhancement, and age of information minimization. Game theory, control theory, machine learning, neural networks, and reinforcement learning techniques can be used to solve these complicated problems. In \cite{qiu2019deep} authors jointly consider the networking, caching and computing problem in hybrid satellite systems. Deep Q-learning algorithm is designed to solve this extremely complex problem.

The SGC and GSC links have strong direct paths and the channel is generally modeled as Additive White Gaussian Noise or Ricean. Due to small channel fluctuations, terrestrial users close to each other cannot be separated through beamforming. In \cite{christopoulos2015multicast}, multi-cast, multi-group beamforming is developed by dividing users spread over large geographical area into multiple groups. Same symbol is transmitted by the DSSN satellite to each group, and digital beamforming techniques are employed for users within each group. Beamforming vectors of DSSN satellite and MTCS can also be jointly optimized for further performance improvements. Overall, the development of centralized and distributed algorithms with different resource availability and CSI considerations in DSSN remain interesting research directions.

\section{Performance Improvements with DSSN: A Case Study}
\label{sat:example}

\begin{figure}[tpb]
\centering
 \includegraphics[width=0.45\textwidth]{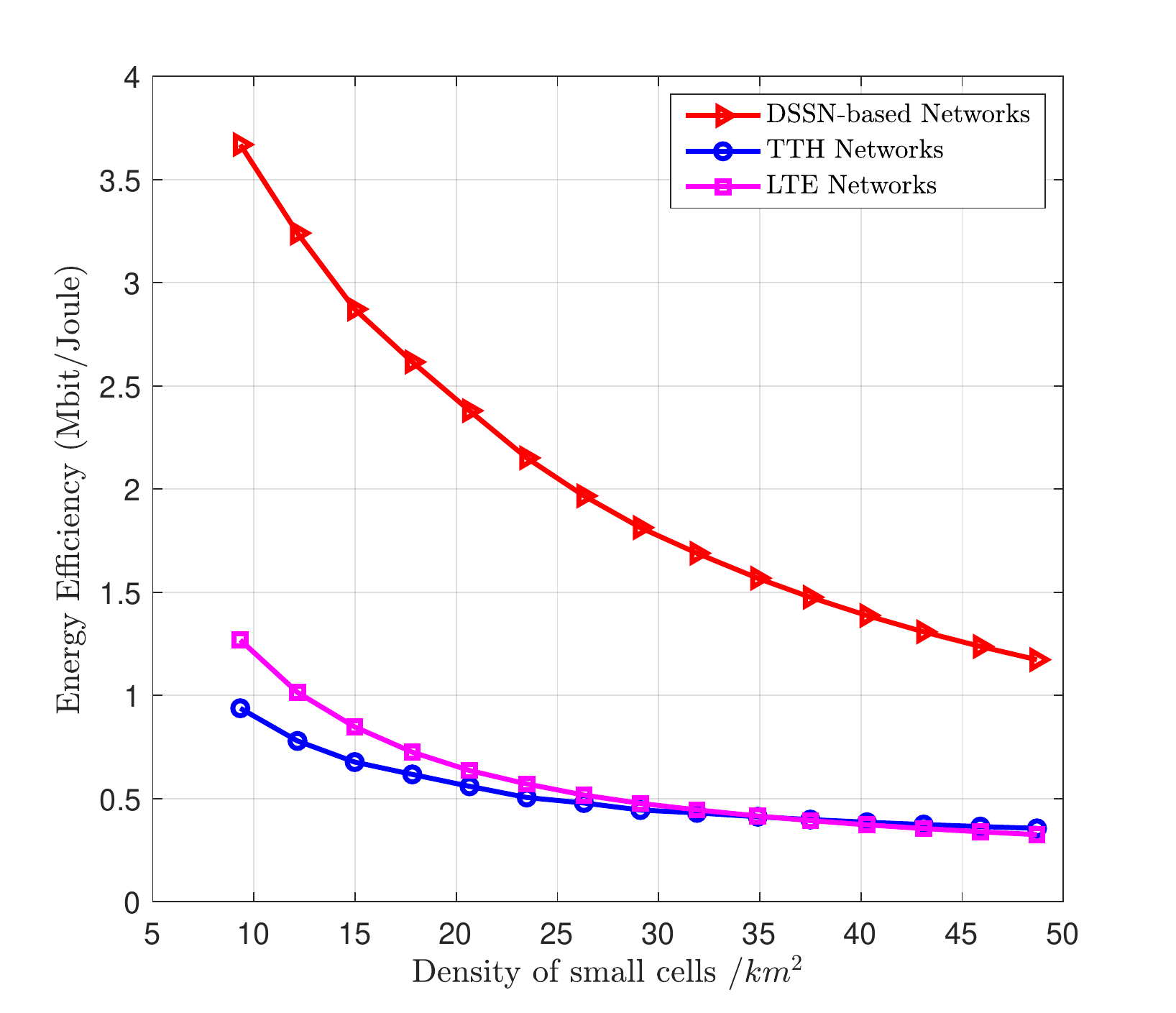} \vspace{-0.2cm}
\caption{Energy efficiency versus density of small cells.}
\label{fig:ee}
\vspace{-0.4cm}
\end{figure}
In this section, we illustrate the performance of energy efficiency versus the density of small cells under different network architectures, which are elaborated in the following:
\begin{itemize}
  \item \textbf{DSSN-based mobile terrestrial networks}: We adopt the high-frequency band (i.e., the Ka-band) to communication, where user devices do not support direct communications, and each user needs the aid of ground infrastructure to access the network via the terrestrial-satellite terminal (TST). This specific scenario corresponds to SGC architecture 2 as given in Table~\ref{sat:arc}. %Fig. \ref{fig:sgc}.
  \item \textbf{Traditional Terrestrial Heterogeneous (TTH) networks}: All small cells are traditionally backhauled via wireless or wired links with limited capacity \cite{di2019ultra}.
  \item \textbf{Long Term Evolution (LTE) networks}: All users access to small cells by the Frequency Division Duplexing (TDD) mode operating at 1800MHz.
\end{itemize}

In our simulations, we set the radius of the macro cell as 1km and each small cell consists of 10 users for all considered scenarios. For the DSSN-based network, the number of TST is 10 for each small cell, and the number of LEO satellites is 8 for this macro cell. The heights of satellites is set as 600km. These simulation settings have been used in several existing papers on DSSN. The power consumption  of each user and small BS are set as 0.2W and 50W, and the maximum transmit power of each TST is 2W. The small scale fading over Ka-band is modeled as Rician fading, and the communication bandwidth is set as 400MHz. Most simulation parameters for TTH and LTE are set based on the 3GPP LTE specifications.

Fig. \ref{fig:ee} shows that the energy efficiency (Mbits/Joule) of all considered network architectures suffers a downward trend with the increase of the density of small cells, since the increase of sum rates in this macro cell is slower than the small cell static power increase. However, the performance of DSSN-based networks has much better energy efficiency than that of TTH and LTE networks. This improvement is due to two reasons: i) DSSN powered by solar energy consume less earth power than the latter two networks and ii) DSSN provide a larger backhaul capacity than TTH and LTE networks, which makes more users successfully access the network, resulting in a higher sum rate. The impact of higher sum rate at lower power consumption therefore translates into a significantly higher energy efficiency for DSSN.

\section{conclusion}
\label{sat:conc}
In this paper, we discussed DSSN benefits for MTCS. Large number of small LEO satellites with limited resources and harsh space environment make DSSN design challenging. It is therefore important to understand DSSN design components and suitable technologies to achieve the requirements of MTCS. We identified DSSN technologies, several open research problems, and some challenges associated with the greater integration of DSSN in MTCS.

With SpaceX already deploying 60 Starlink satellites in 2019, DSSN is becoming realistic and popular in very near future. However, significantly novel research efforts in the domains of smart steerable antenna design, NOMA, energy harvesting and optimization, routing and networking protocols, re-configurability, data caching, and resource optimization techniques are still needed. In particular, networking in space is a considerable challenge along with the maintenance and re-configurability issues of satellites in DSSN.

\bibliographystyle{ieeetran}   %% style of IEEE transactions
\bibliography{references} %% Name of your bibtex file

\end{document}